# Exploring Ancient Architectural Designs with Cellular Automata

Hokky Situngkir
[hokky.situngkir@surya.ac.id]
Dept. Computational Sociology, Bandung Fe Institute
Center for Complexity Studies in Surya University

**Abstract**
The paper discusses the utilization of three-dimensional cellular automata employing the two-dimensional totalistic cellular automata to simulate how simple rules could emerge a highly complex architectural designs of some Indonesian heritages. A detailed discussion is brought to see the simple rules applied in Borobudur Temple, the largest ancient Buddhist temple in the country with very complex detailed designs within. The simulation confirms some previous findings related to measurement of the temple as well as some other ancient buildings in Indonesia. This happens to open further exploitation of the explanatory power presented by cellular automata for complex architectural designs built by civilization not having any supporting sophisticated tools, even standard measurement systems.

*In architecture,*
*the pride of man,*
*his triumph over gravitation,*
*his will to power,*
*assume a visible form.*
*Architecture is a sort of oratory of power by means of forms.*
*- Friedrich Nietzsche-*

**1. Traditional Architecture: From Simple to Complex**
There are no evidence that ancient Indonesian society had a metric standard for the precisions and geometry on which they built the civilian constructions. Yet, ruins of buildings and artifacts expressing complex mega-constructions are there, spreading throughout the archipelago. As proposed by Atmadi (1988) and computationally elaborated in Situngkir (2010), we can see that the Borobudur, the biggest Buddhist temple and heritage from ancient Indonesian civilization, use some sort of ratio conjectured to be used by the architect of the temple in overcoming the lacking standard of measurement. In the latter, the algorithmically built temple has fractal geometry with dimension ± *2.3252*. The self-similarity of the building is shown to be emerged from the way of building stupa, Buddhist's relic as the basic shape from which the Borobudur was built. Apparently, the shape of the stupa, with the hypothesized ratio applied, is obvious in a lot of sizes, from a small 3-dimensional ornaments to the shape of the temple itself.

A challenging task is, nonetheless, trying to have answers on how the ancient could possibly built such complex mega-structures with such "simple" tools, devices, and standard measurements. This is the motivation on bringing the report. In this case, we propose the concept of cellular automata – a computational model that has been so famous explaining how complex structures emerged from few simple rules (Wolfram, 2002).

Some issues related to the explanation on aesthetic artifacts scientifically are discussed in Barrow (2003). While complexity studies have introduce many possibilities in arts and aesthetics, it also happens to open possibilities to learn a lot of traditional artifacts that was made by cultures – alternative to conventional geometry and art studies (Situngkir, 2005). It has been reported that some traditional settlements in Africa also accidentally follows fractal structure (Eglash, 1999) as they grow organically from time to time. The studies into Indonesian traditional craft, batik, also explain the fractal geometry in it (Situngkir, 2009).

Cellular automata is very interesting computational model showing how the complexity of natural phenomena came from very simple rules run iteratively. The simplicity that is dependent on the local interconnectivity of cellular automata has been recognized to be good candidate for efficient, yet massive microphysical architectural implementation (Margolus, 1995). In advance, some fundamental issues on algorithmic architecture, including those with cellular automata, are elaborated in detail by Terzidis (2006). Modern contemporary life has witnessed how cellular automata become an interesting source of inspiration for modern architectural designs as generative art (Krawczyk, 2002). The latest has been understood to be the kind of generative architecture, a part of the more general generative arts. Conjectures to the employment of cellular automata in contemporary architectural designs have been widely developed and improved recently Cawley & Wolfram (2010).

This paper presents the exploratory study of cellular automata into general shape of Indonesian ancient buildings, temples, that happen to be mega-structures with precisions, yet were built without any complicated tools, devices, and geometry as we know today. The idea is simply from the realizations that the complex architectural designs were emerged from simple local rules in the micro-view.

## 2. Cellular Automata as Exploratory Tool

Our hypothesis is based upon the 3-dimensional cellular automata constructed within the totalistic 2-dimensional cellular automata with 9 neighbors (Packard & Wolfram, 1985). The two dimensional cellular automata consists of regular lattices of sites, where sites have *k*-possible states,. The dynamicity of the system occurs by updating the sites in discrete time steps according to certain rules ($\phi$). The value of the site in particular position, $s_{ijk}$, follows,

$$s_{ij}^{t+1} = \phi\left[ s_{i-1,j-1}^t, s_{i,j-1}^t, s_{i+1,j-1}^t, s_{i-1,j}^t, s_{i+1,j}^t, s_{i+1,j-1}^t, s_{i+1,j}^t, s_{i+1,j+1}^t \right] \tag{1}$$

where in the totalistic rules, $\phi$ is simply summing up the values of all the neighbors,

$$s_{ij}^{t+1} = f\left[ s_{ij}^t + s_{i-1,j-1}^t + s_{i,j-1}^t + s_{i+1,j-1}^t + s_{i-1,j}^t + s_{i+1,j}^t + s_{i+1,j-1}^t + s_{i+1,j}^t + s_{i+1,j+1}^t \right] \tag{2}$$

From the rules, the number of possible cellular automata rules are,

$$N = k^9(k-1)k \tag{3}$$

and for only two possible states, as we are utilizing in the paper, we have $N = 1024$ possible rules. Thus we can write the function for *n*-th rules of 9-neighbors totalistic cellular automata as,

$$C = \sum_n f(n)k^n \tag{4}$$

The phenomenological study observe all possible rules as each grows as the lower plan of architectural designs to the top. This is expected to bring us to the conjectures of explanation and reflection the computational generic process in the architecture. Our study tries to show how the simple rules of cellular automata apply from the lowest to the top of the building.

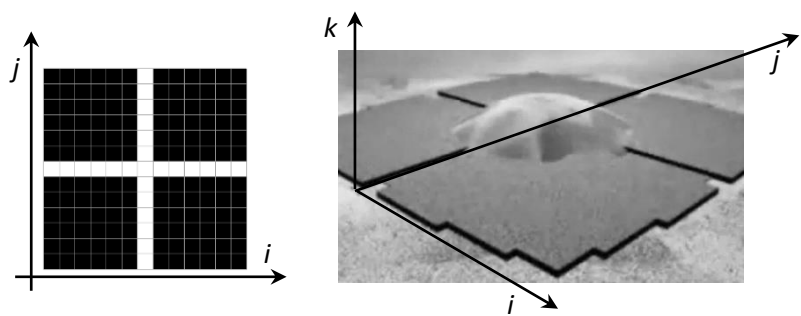

**Figure 1**
The lower ground of Borobudur (*right*) used as initial condition in the cellular automata (*left*)

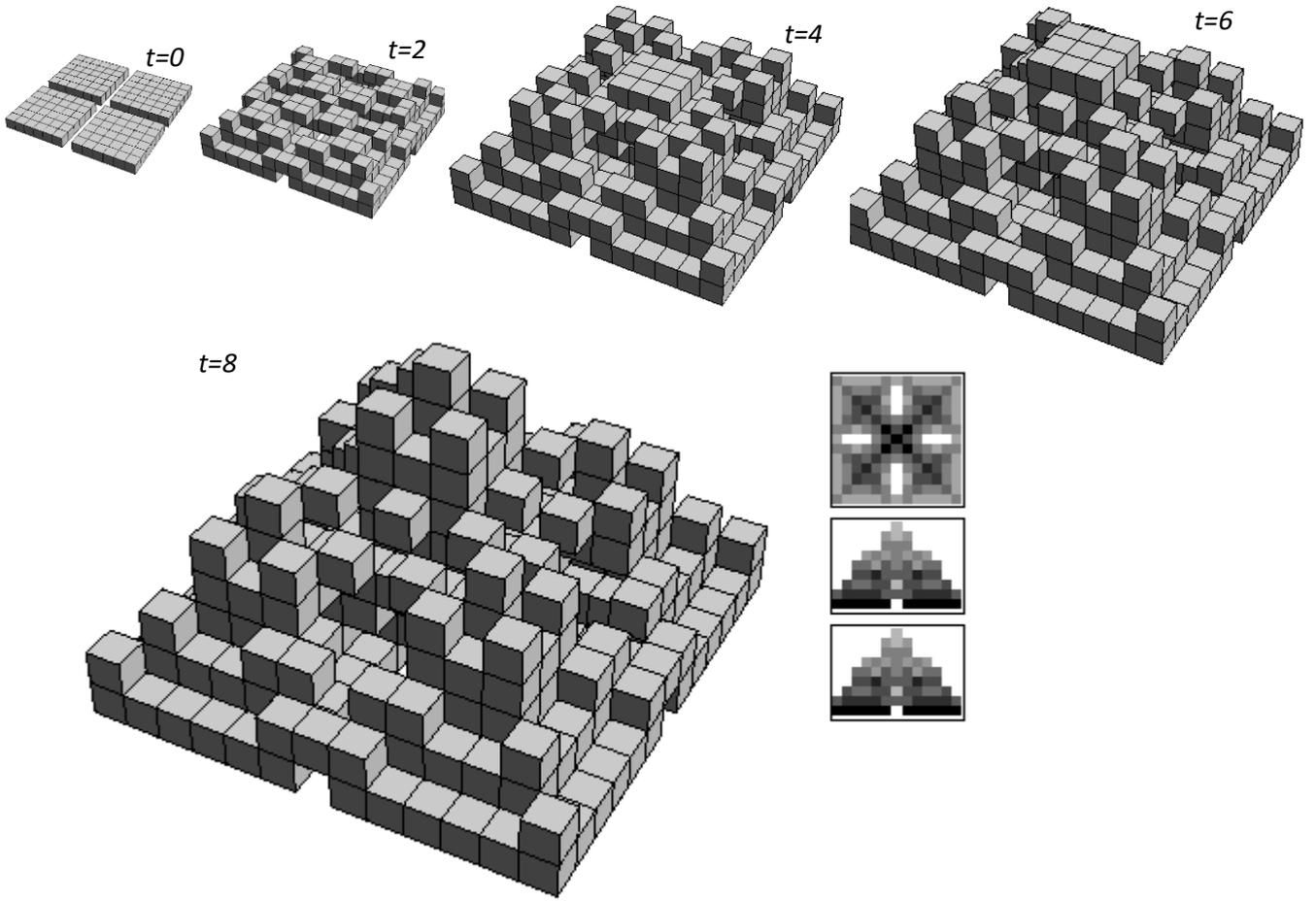

**Figure 2**
Generic steps of cellular automata rule *C=816* the shape that is hypothetically corresponded to Borobudur Temple in Indonesia.

We use the shape of lower ground of the architectural structure as the initial condition of the cellular automata. The rules of the totalistic 9 neighbors 2 dimensional cellular automata are then implemented generically to upper floor sequentially. In three-dimensional space, we re-write eq. (2) into,

$$s_{i,j,k} = f\left[ s_{i,j,k} + s_{i-1,j-1,k} + s_{i,j-1,k} + s_{i+1,j-1,k} + s_{i-1,j,k} + s_{i+1,j,k} + s_{i+1,j-1,k} + s_{i+1,j,k} + s_{i+1,j+1,k} \right] \quad (5)$$

It is worth to note, that in our exploration, as the iteration steps, the horizontal length of each layers are delimited to the size of the initial layer (Fig. 1) since the subsequent layers can always larger, which is, roughly speaking, not realistic in the observed architecture designs.

In advance, we have the 3-dimensional form that would be able to be compared with the scrutinized architectural designs. In this case, we show that the cellular automata can be utilized as an explanatory tool for the ancient complex buildings, as well as the source of inspiration to the modern architectural designs.

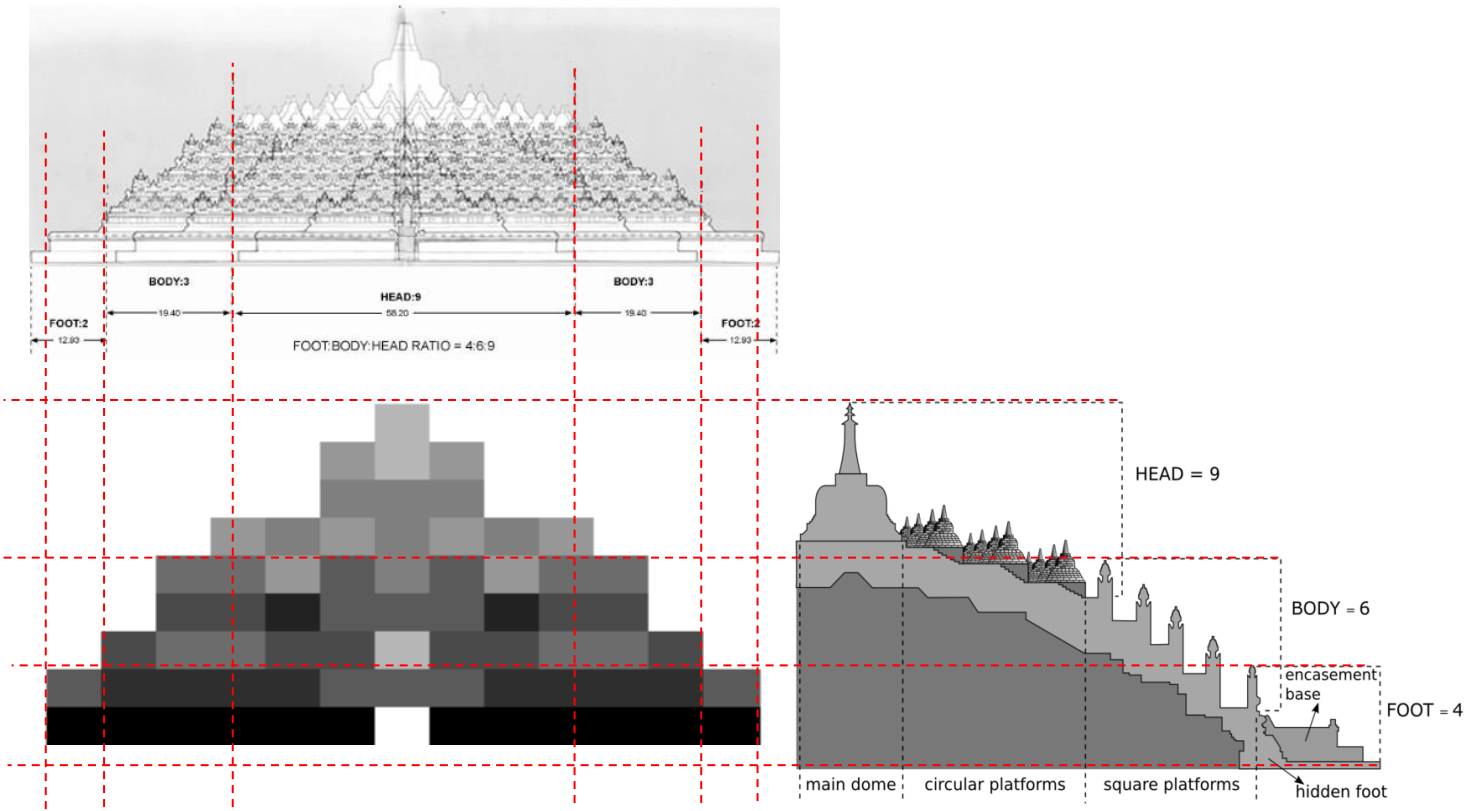

**Figure 3**
The emerged vertical plan of the totalistic cellular automata rule *C=816* is compared with those of the sizes in Borobudur Temple as recorded in Atmadi (1998) & calculated in Situngkir (2010).

We observe the interesting shape of Indonesian Borobudur Temple and compare to those of the emerged three-dimensional patterns. Here, we use the lower ground of the temple as the initial condition as shown in figure 1, and found that the rule *C=816* is similar to the shape of the temple. The evolving steps are shown in figure 2. A more detail comparison is presented in figure 3. Here the rule 4:6:9 as hypothesized by Atmadi (1998) can also be seen in the vertical plan of the emerged 3-dimensional pattern. By using this perspective, we could say that the rule 4:6:9 as shown in the emerged big picture of the building, instead of being the basic plans of the building. However, as shown in Situngkir (2010), small stupas might be seen as the basic shape acting as the cuboids utilized in the cellular automata. This is apparent in the building reflecting the self-similarity and the signature of the fractal geometry in the architectural design, in general.

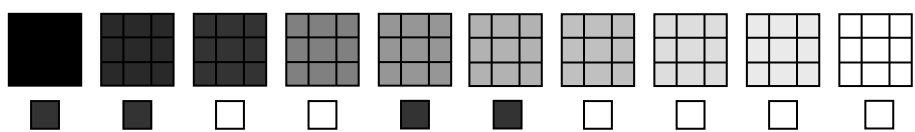

**Figure 4**
The 9-neighbors 2-dimensional totalistic cellular automata rule *C=816*

| Architecture | | CA Rule | Top/Side Plan |
|---|---|---|---|
| **Prambanan, Central Java** – the largest Hindu temple in Indonesia, from the dynasty of Sanjaya (850 AD) | 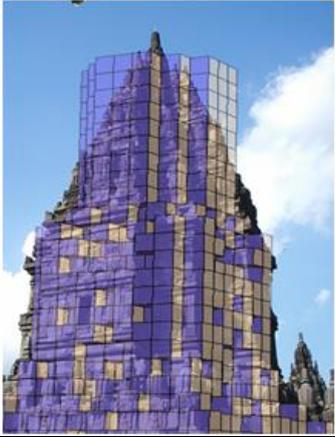 | 944 | 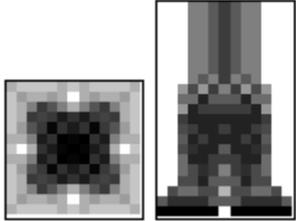 |
| **Sukuh Temple, Western Slope Mt. Lawu, Central Java**, a Javanese Hindu temple from Majapahit Kingdom (1437 AD). | 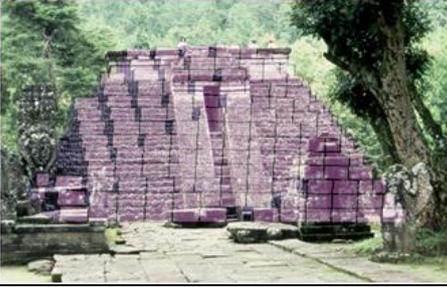 | 960 | 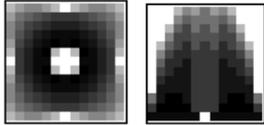 |
| **Mendut Temple, Central Java**, a Buddhist temple located about 3 km from Borobudur, from Syailendra Dynasty (824 AD) | 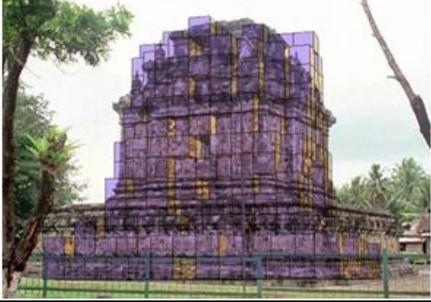 | 688 | 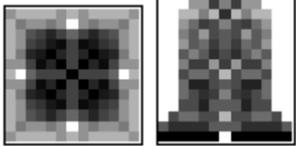 |
| **Cangkuang Temple, Garut, West Java**, a Hindu temple, approximately from 8[th] century. | 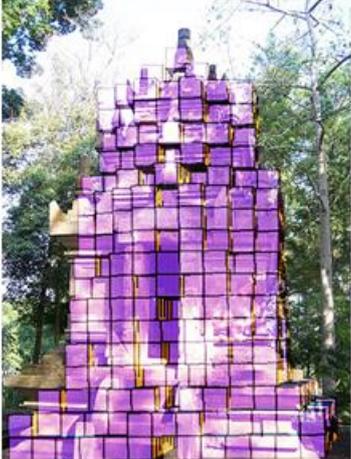 | 688 | 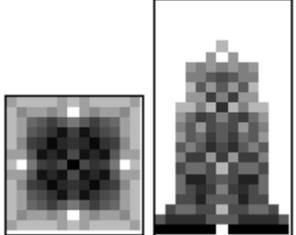 |

**Figure 5**
Various temples from Indonesian ancient times that can be observed as complex structures of yielded with rules of totalistic cellular automata.

The rule *C=816* is represented to be in binary form of $1100110000_2$. This reflects the enumeration of the updating rule in particular the totalistic cellular automata, $C_{binary} = \{1,1,0,0,1,1,0,0,0,0\}$. We have only two colors used here, $k=2$, and the rule is illustrated in figure 4. It is shown that the emerged pattern as seen in figure 2 comes from the simple rules of which four possible additions of cuboids as denoted in equation (2). The rule is expressing that the addition of small cubes in the next layer is conducted when there are 4, 5, 8, or whole 9 neighboring sites are covered by cubes in the respective previous layer,

$$s_{i,j,k+1} = \begin{cases} 1, & 5 \leq \sum_{i,j} s_{i,j,k} \leq 6 \land 9 \leq \sum_{i,j} s_{i,j,k} \leq 10 \\ 0, & otherwise \end{cases} \quad (6)$$

The placement of new elementary block in a particular location is dependent to the existence of the elements surrounding the same location in the previous layer. Obviously, this is a very simple rule that is capable to emerge such complex architectural designs like Borobudur.

Series of experiments are thus conducted into other ancient buildings in Indonesia. Interestingly, lots of them are have interesting similarity in their shape to particular cellular automata emerged from certain totalistic rules as shown in figure 5. Nonetheless, it is also possible to have alternating views on rules to fit the shape of the ancient buildings. For instance, it is not implausible to think that a building might be built with two or more rules applied. In designing a complex buildings like temples, one might implement different rules in constructing the building and the roof if it.

## 3. Discussions

Wolfram (1984) showed four qualitative classes of cellular automata related to patterns that would turn out to be the limit points of all configurations after some periods. The first class is the when all the sites eventually have the same value after a relatively short transient period (limit points). The second class is related to the homogenous state or becomes some stable and oscillating patterns (limit cycles). The third one is when the cellular automata exhibit chaotic behavior or an apparent unpredictability pattern. The fourth class is associated to complex behavior as the cellular automata exhibits long transients and propagating patterns. When organic and natural systems seem to favor the fourth class to be the most suitable cellular automata, generic process of architecture using cellular automata seemed to have class I, the limit points.

The table in fig. 5, as well as fig. 2, show the plans of cellular automata that is fit in shape to the ancient architectural designs. It seems that most ancient architectural designs to be modeled with 2-dimensional totalistic rule cellular automata are in the class I. Indirectly, this is related to report by Wolfram (1984) that as class IV cellular automata appear to be much less common in two dimension than in one dimension, the 2-dimensional cellular automata are overwhelmed with the class III ones. The rare special totalistic rule for the cellular automata of class IV is recognized in the variants of the famous "Game of Life". The buildings are not categorized to be in limit cycles either, for the plausibility that ancient buildings do not to have the cyclic designs.

Architectural designs are apparently associated to orders. For class I cellular automata, their complexity is not supposed to exhibit such chaotic patterns nor propagating ones, in general. This happens to be upper boundaries of the cellular automata. However, there should also such transient period, not too short, for the structures of ancient architectural designs to be fitted in.

It is almost definite that how we approach the ancient architectural designs in this way brings us only to confirmed hypothetical understandings on how such complex structures can be erected among the civilizations lack of standard measurement system and simple devices. There is no evidence that the respective civilization has recognized some "analogue" version of the computational method we have here. Yet, this approach has shown another explanatory power of the cellular automata: how simple social life can have such complex and (some of them) magnificent architectural designs.

## 4. Concluding Remarks

Observing the complexity of architectures from ancient social life, like temples, that are found a lot in Indonesian archipelago, is often bringing question on how such simple civilization could erect them regarding to the known technical simplicity they had. This question is sometimes followed by appearing mysteries related to detail within them as our modern eyes scrutinize each of them. The simple method emerging complex patterns as shown in cellular automata is hypothesized to be able answering the question.

The paper reports the utilization of three-dimensional forms emerged by the two-dimensional totalistic cellular automata with some modifications related to the delimitation of the growing sites horizontally. The emerging 3-dimensional forms are compared with some ancient temples in Indonesia. The more detailed observation of the emerged 3-dimensional shape gives more interesting result, related to the ratio 4:6:9 discovered in multi-scaled measurements of Borobudur Temple previously, which is emerged from more elementary and simple rules of particular cellular automata. Furthermore, our discussions conclude some characteristics of the utilized cellular automata used in the observation.

This observation confirms the explanatory power of cellular automata to ancient architectures. This is a supplementary to the widely recognized exploratory power of cellular automata as inspiration to the modern and contemporary architectural designs.

**Acknowledgement**
Author thanks Surya Research International for support in which period the research is conducted and people at Bandung Fe Institute for discussions of the rough version of the paper.